\documentclass{ws-procs9x6}

\begin{document}

\title{SELF-ORGANIZING MAPS AND PARTON DISTRIBUTION FUNCTIONS}

\author{K. HOLCOMB}

\address{%
University of Virginia - UVACSE \\  
112, Albert H. Small Building, Charlottesville, Virginia 22904 - USA \\ 
E-mail: \email{kholcomb@virginia.edu}} 

\author{S. LIUTI$^1$, D.Z. PERRY$^2$}

\address{%
University of Virginia - Physics Department \\
382, McCormick Rd., Charlottesville, Virginia 22904 - USA \\ 
$^1$ E-mail: \email{sl4y@virginia.edu} 
$^2$ E-mail: \email{dzp3h@virginia.edu}}

\begin{abstract}
We present a new method to extract parton distribution functions from high energy experimental data  based on a specific type of neural networks, the Self-Organizing Maps. We illustrate the features of our new procedure that are particularly useful for an anaysis directed at extracting generalized parton distributions from data.
We show quantitative results of our initial analysis of the  parton distribution
functions from inclusive deep inelastic scattering. 

\end{abstract}


\bodymatter

\section{Introduction}\label{sec1}
High energy processes can be described within Quantum Chromodynamics (QCD) through its factorization properties. These allow us to write the cross sections by separating the hard part, which is the well defined interaction of a quark with {\it e.g.} a virtual photon, from the soft part, described in terms of Parton Distributions Functions (PDFs). The latter cannot be calculated from first principles.
Information about the PDFs can only be obtained directly from experiment.  Several collaborative groups  in the past decades have sought to use fitting techniques to study the PDFs behavior by devising models/analytic forms  for the various PDFs, $f_i(x,Q^2)$, $i=u,d,s.,c,b,t,g$, at light cone momentum fraction $x$, and scale $Q^2$, whose parameters are constrained by experiment.

More recently, new types of partonic distributions that improve considerably our  ability to study the nucleon structure and its partons dynamics became accessible experimentally. The new distributions extend, in different ways, the concept of PDFs to  observables that can  measure the spin, spatial and momentum correlations in hadrons. These are the Transverse Momentum Distributions, TMDs Ref.\refcite{MulTan}, and the Generalized Parton Distributions, GPDs (see review in Ref.\refcite{BelRad}). For GPDs and TMDs one increases the range of kinematical variables that need to be measured by introducing both transverse momenta, and the Fourier conjugates of the partonic spatial degrees of freedom. 

While for the PDFs one faces the problem of considering different results/extractions that vary significantly from group to group, lacking a well defined theory to constrain them (for a review of recent results see {\it e.g.} Refs.\refcite{Dittmar,PDF_WG}), for TMDs and more crucially for GPDs, standard global analyses are doubtful due to the increased number of variables compared to the scarcity of the experimental data
(see {\it e.g.}  Ref.\refcite{GuiMou} for a state of the art analysis). 

A relatively new approach to PDFs fitting is the one proposed by
the NNPDF collaboration Ref.\refcite{NNPDF},
who have replaced the standard analytic forms of PDFs with 
a more complex  Neural Network (NN) solution. 
The estimated uncertainties for NNPDF fits are larger than those of global
fits, possibly indicating that the global fit uncertainties might be underestimated.
In Ref.\refcite{Carnahan} a criticism was put forward about relying on purely automated fitting procedures such as the ones used by NNPDF. A new specific type of neural network, the Self-Organizing Map (SOM), was proposed. 
The main point is that since for NNPDFs the effect of modifying individual
NN parameters is unknown, the result might not be under control
in the extrapolation region, or in between the data points if the data are 
sparse. 
This issue is even more important when extending the fitting procedures to a wider set of semi-inclusive and exclusive observables. We therefore pushed forward with the SOM method, and we improved the preliminary work in Ref.\refcite{Carnahan} by restructuring the original code in such a way that 
on one side, a fully quantitative error analysis can be implemented, and on the other we now have sufficient flexibility to allow for analyses of different observables, including  the matrix elements for deeply virtual exclusive and semi-inclusive processes.
Our first quantitative results for the unpolarized case using Next-to-Leading-Order (NLO) perturbative QCD in the $\overline{MS}$ scheme were presented in Ref.\refcite{Perry_dis10}. In this workshop we also discuss a possible procedure to extend SOMs to the extraction of GPDs.

\section{Self-Organizing Maps}
\label{sec2} 

Recent years have seen an outstanding growth in the usage of neural networks as  paradigms for computational methods. SOMs are a type of neural network developed by T. Kohonen in the '80s Ref.\refcite{Kohonen} based on a topological mapping of the external environment  onto the brain's internal neural connections. 

\begin{figure}
\centering
\centerline{\includegraphics[width=0.35\textwidth]{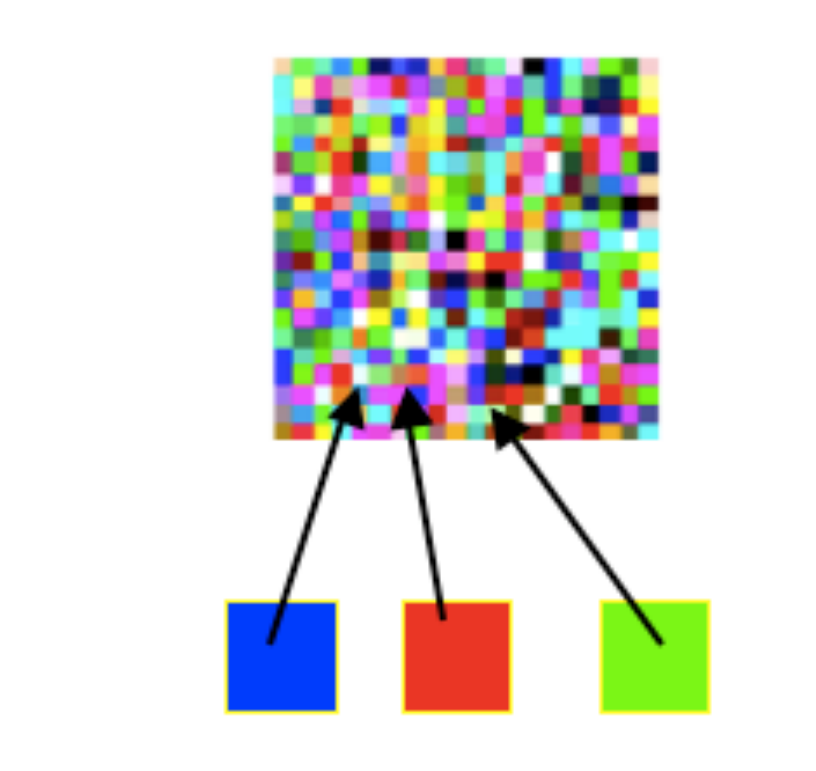}}
\caption{A simple illustration of a $20\times 20$ square map, and its input signals represented by colors. The arrows indicate the signal to cell matching procedure. In our more realistic situation, the map and signals will constitute of either PDFs or DIS structure functions.}
	\label{fig0}
\end{figure}

In SOMs the  nodes/neurons -- map cells -- are tuned to a set of input signals/data/samples according to a form of adaptation. The various nodes form a topologically ordered map during the learning process. This feature constitutes perhaps the main strength of SOMs in that it allows one to obtain a simplified two-dimensional representation of complex data otherwise depending on a "hard to control''  set of parameters (see Figure \ref{fig0}). In this respect SOMs can be considered as a non linear form of principal component analysis which is often used to analyze  highly dimensional data.

Another aspect that sets SOMs apart from other NNs  is their learning process. SOMs learn via {\em unsupervised learning} whereas the learning 
of generic artificial NNs is {\em supervised}. 
In supervised learning a set of examples is given, and the goal is to force the data to match the examples as closely as possible.  A cost function is defined that measures the importance to miss or detect the correct result. During the learning process the cost function is minimized. In unsupervised learning the cost function is minimized without introducing a definite set of examples, but just by similarity relations, or  by finding how the data cluster or self-organize.  Theoretical studies of the mechanism for map evolutions are in progress. In phenomenology, many new 
uses, one of which  described below, might derive from this property.    

SOMs are built as two dimensional arrays  whose cells get sensitized/tuned to a specific set of input  signals according to a given order.  To illustrate this in Fig.\ref{fig0} we show the simple example of a "color map", where the input signals are represented by the three colors to be associated with each one of the map cells. 
 
The SOM algorithm consists of three stages: {\it i)} Initialization; {\it ii)} Training; {\it iii)} Mapping. 
 
During the initialization procedure weight vectors of dimension $n$ are associated to each cell $i$:
\[ V_i = [v_i^{(1)}, ..., v_i^{(n)} ]  \] 
$V_i$ are given spatial coordinates, {\it i.e.} one defines the geometry/topology of a 2D map
that gets populated randomly with $V_i$. 
For the training, a set of input data
\[ \xi = [\xi_i^{(1)}, ..., \xi_i^{(n)} ] , \] 
(isomorphic to $V_i$) is then presented  to $V_i$, or compared  via a "similarity metric" that we choose to be
\[ L_2(x,y) = \sum_{i=1,2} \sqrt{  x^2_i - y^2_i  } \]
The most similar weight vector is the Best Matching Unit (BMU). 

SOMs are based on  unsupervised and "competitive" learning.
This means that the cells that are closest to the BMU activate each other in order to "learn" from $\xi$.  Practically, they adjust their
values according to
\begin{equation}
V_i(n+1) = V_i(n) +  h_{ci}(n) [\xi(n) - V_i(n)],
\label{learn_1}
\end{equation}
where $n$ is the iteration number, and $h_{ci(n)}$ is the "neighborhood function" defining a radius on the map which decreases with both $n$, and
the distance between the BMU and node $i$. In our case we use square maps of size $L_{MAP}$, and 
\begin{equation}
h_{ci}(n) = 1.5 \left(\frac{n_{train} -n}{n_{train}} \right) L_{MAP}
\label{learn_2}
\end{equation}
where $n_{train} $ is the number of iterations.
At the end of a properly trained SOM, cells  that are topologically close to 
each other will contain data which are similar to each other.
In the final phase the actual data are distributed on the map and 
clusters emerge.   (Note that the specific location of the clusters on
the map is not relevant and will not necessarily be the same from one run to another;
only the clustering properties are important.)

Since each map vector now
represent a class of similar objects, the SOM
is an ideal tool to
visualize high-dimensional data, by projecting it onto a low-dimensional map
clustered according to some desired similarity feature. 

\section{SOMPDF Parametrization} 
\label{sec3}
In Ref.Ref.\refcite{Carnahan}  a SOM algorithm was constructed which, together with a Genetic Algorithm (GA) was applied to PDF fitting. In this initial work it was proven that the SOM method works well as a minimization technique, in that it was shown that the $\chi^2/$d.o.f. decreased towards unity with each GA iteration. However problems connected with the smoothing of the functions remained, related to the stochastic procedure used to generate them and to initialize and train the map. These prevented a fully quantitative analysis of the data, including uncertainty evaluations.

A solution to this problem was obtained in the present work Ref.\refcite{Perry_dis10,progress} by developing a new version of the SOMPDF code where several important improvements were introduced.  By writing it in a single compiled language -- fortran 95 -- the code was made more flexible and 
faster since a parallel (MPI) version was easily implemented. As explained in more detail below, the increased flexibility and speed allow us to perform now 
random variations on the parameters of the various input PDFs, instead of on the PDF values themselves,  thus providing continuous solutions which are amenable to standard error analyses. 

Our ultimate goal is to provide a procedure to extract the GPDs, and their related observables, including {\it e.g.} the spatial partonic d.o.f. from experimental data. We therefore added flexibility in the code's inputs so as to accommodate the additional kinematical variables and parameters.

We now describe the main fitting procedure, originally applied to PDFs. 
This first phase of our work involves only marginally the clustering and visualization properties of SOMs.  For PDFs this is, in fact, barely needed since the inclusive scattering databases provide a rather large kinematical coverage, and only two kinematical variables are necessary to describe the data. SOMs are used mostly to realize random variations of the PDF curves. 

Our fitting procedure is based on the initialization, training, and  
mapping steps described for a general case in Section \ref{sec2}.
In a nutshell, a set of database/input PDFs is formed by selecting at random from a range of existing PDF functions and varying their parameters. Baryon number and momentum sum rules are imposed at every step. These input PDFs, evolved to NLO to the desired $Q^2$,  are used to initialize the map. A subset of input PDFs is then used to train the map. Notice that this phase of the procedure implies defining specific criteria for selecting the PDFs that will define each subset. Two possible criteria are discussed in Ref.\refcite{Carnahan}. 

\begin{figure}
\centering
\centerline{\includegraphics[width=0.45\textwidth]{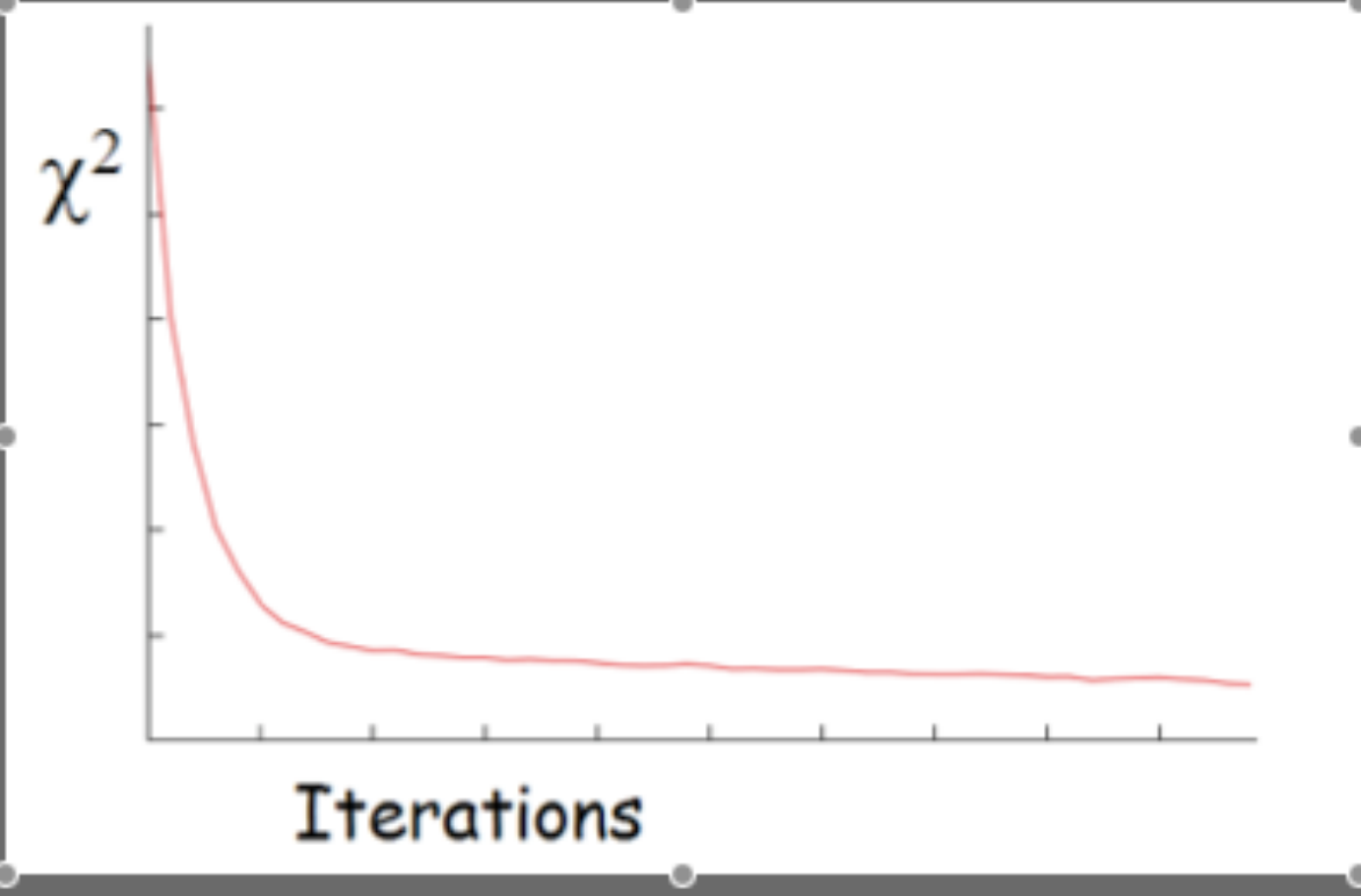}}
\caption{Illustration of the behavior of the $\chi^2$/d.o.f for one of our SOMPDF runs.}
	\label{fig1}
\end{figure}

The similarity is tested by comparing the PDFs, according to Eqs.(\ref{learn_1})  and (\ref{learn_2}) 
at given $x$ and $Q^2$ values. The new map PDFs are obtained by averaging the neighboring PDFs with the BMU PDFs. Once the map is trained a GA is implemented. The $\chi^2$ per input PDFs is calculated with respect to the experimental data. We then take a subset of these functions with the best  $\chi^2$, and use them as seeds to form a new set of input PDFs. We train the map with the new set of input PDFs and repeat the process. The $\chi^2$ was found to decrease monotonically towards $\chi^2=1$ with every GA iteration (Fig.\ref{fig1}).  Our stopping criterion is established when the $\chi^2$ stops varying -- its curve flattens.
\begin{figure}
\centering
\centerline{\includegraphics[width=0.75\textwidth]{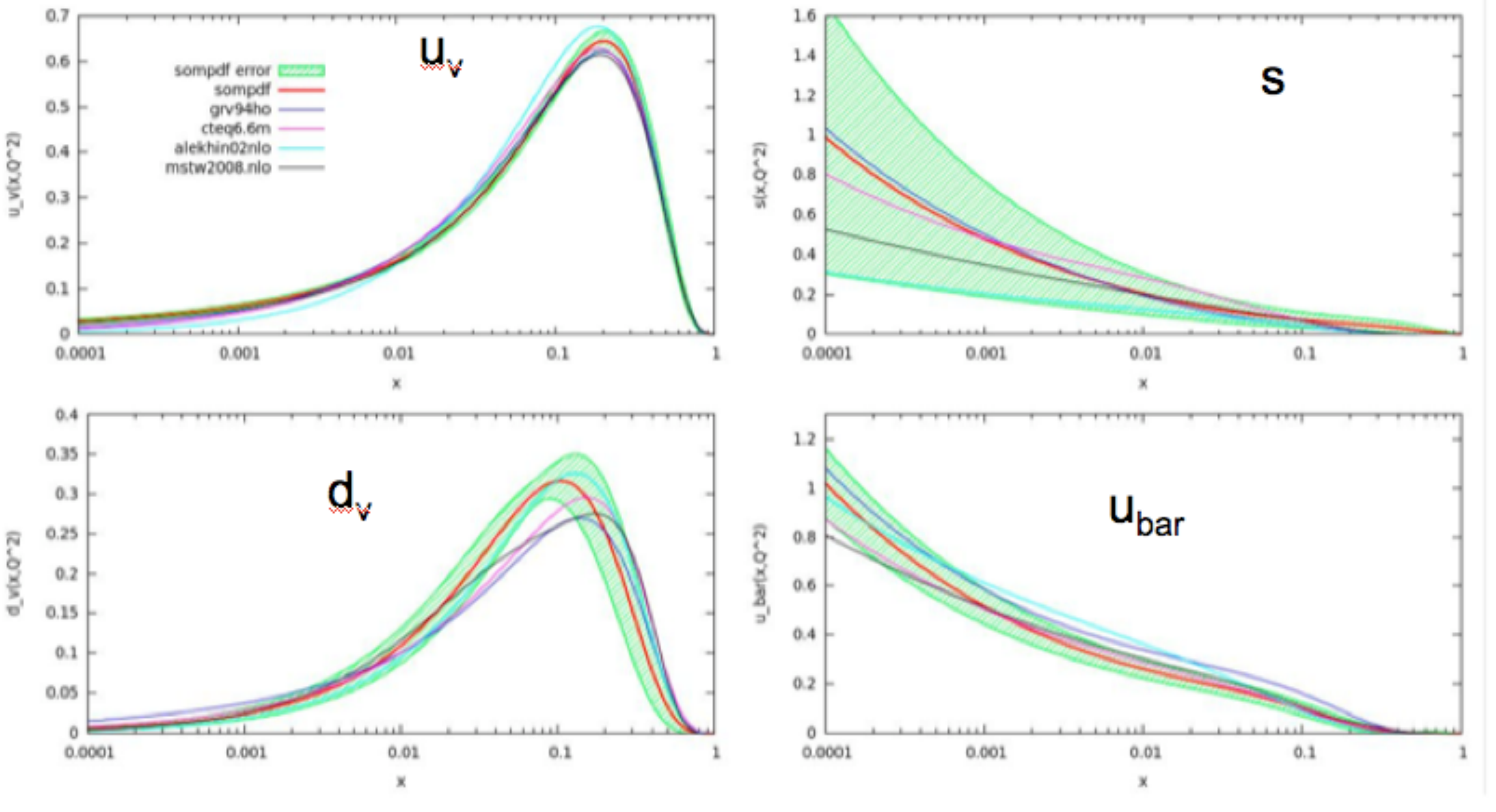}}
\centerline{\includegraphics[width=0.75\textwidth]{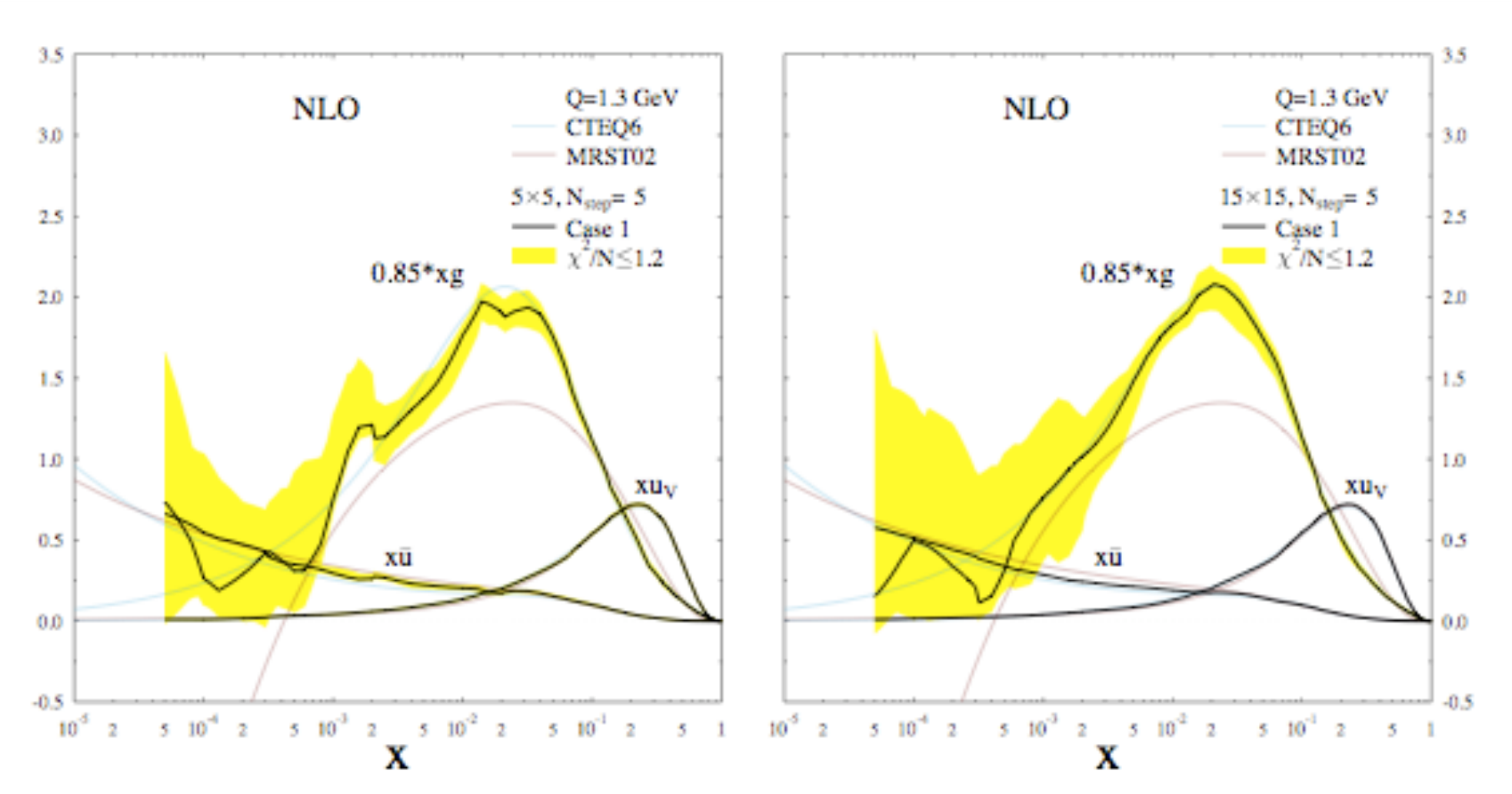}}
\caption{Upper panel: Test results using a $5 \times 5$ map for a set of 43 runs. The panels represent (clockwise from upper left)the $u_v$,
$s$, $d_v$, $\bar{u}$ distributions, respectively, at $Q^2 = 7.5$ GeV$^2$. The shaded areas are our results including the error analysis
outlined in the text. For comparison we show also results from several other parametrizations.
Lower panel: $u_v$, $\bar{u}$ and gluon distributions at $Q^2 = 1.3$ GeV$^2$, along with their variations obtained for $\chi^2/$ d.o.f. $<1.2$, from our previous SOMPDF analysis. Left: $5 \times 5$ map, right: $15\times 15$ map. These figures are shown to illustrate the improvement in the curves smoothing attained in our new procedure (adapted from Ref.\refcite{Carnahan}).}
	\label{fig2}
\end{figure}

Our first runs, presented in Figure \ref{fig2}, used a "test" set of data from DIS  consistent with the sets used in Refs.\refcite{NNPDF,CTEQ6,AMP06}.  The data sets chosen were from 
BCDMS, H1, NMC, SLAC and ZEUS (see {\it e.g.} references in Ref.\refcite{NNPDF}). The shaded areas in the figure represent our error bands. In our preliminary runs we defined a statistical error on an ensemble of SOMPDF runs. 

\begin{figure}
\centering
\centerline{\includegraphics[width=0.85\textwidth]{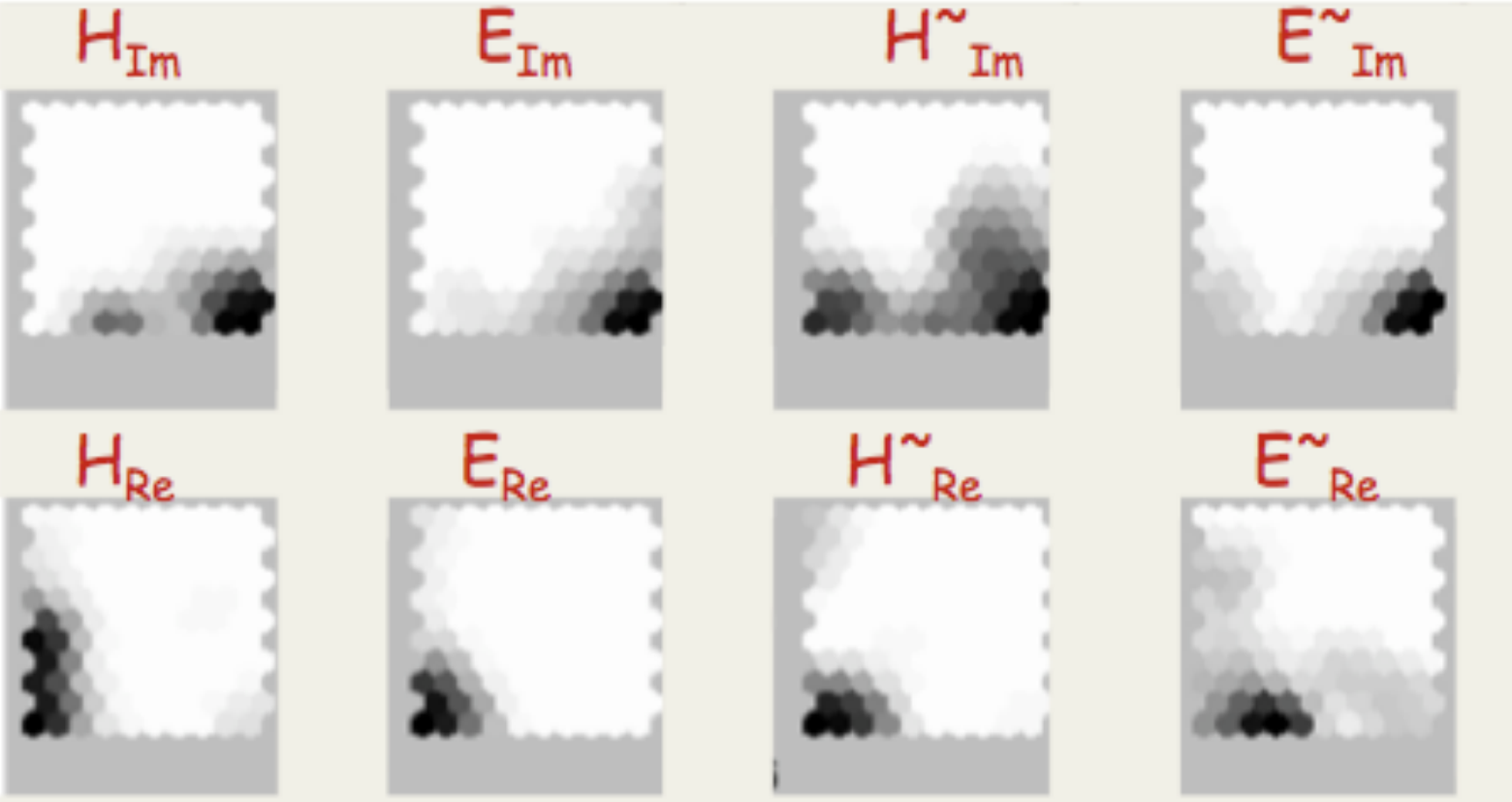}}
\caption{Extension of our analysis to GPDs.} 
	\label{fig3}
\end{figure}

Finally, with a fully working procedure in hand, we comment on our capability to extend the PDF analysis to the GPD case. 
Here one has a total of $8$ observables per parton component, given by the real and imaginary part of the Compton Form Factors (CFFs) containing the corresponding GPDs Ref.\refcite{BelRad}. The number of observables that has been defined {\it e.g.} in Ref.\refcite{GuiMou} is $17$ of  which $6$ appear at Leading Order. An open question is therefore to establish which experiments, observables, and with what precision they can determine the various GPD components. Our analysis will provide a step in this direction in that by applying the clustering properties of the SOMs we will be able to quantitatively discern the various CFFs, that will become "dimensions" in our analysis, and to establish the sensitivity of the different experiments to each one of them. An illustration of how clustering of GPDs on the SOM might be represented is given in Fig.\ref{fig3}. 
More work is in progress Ref.\refcite{progress}. 

\section{Conclusions}
In this work we described a new computational method based on Self-Organizing Maps for parametrizing nucleon PDFs. 
Future developments will also be directed at exploiting the full potential of SOMs that  offer the capability of going beyond a fully automated procedure, by enabling one  to control the fitting procedure at each step. 
The selection of the best PDF candidates for the subsequent iteration could 
then be made based on the user's preferences instead of solely based on the
 $\chi^2$.
Our program can be extended
to multivariable cases such as the Generalized Parton
Distributions where the data are too
sparse for stochastically generated, parameter-free, PDFs.

\vspace{0.3cm} 
We thank the University of Virginia Alliance for Computational Science and Engineering for computer time, and the HPC group at Jefferson Lab, in particular David Richards and Chip Watson, for
allotting us space on their clusters. This work is partially supported by the U.S. Department
of Energy grant DE-FG02-01ER4120 (S.L and D.Z.P.).


\begin{thebibliography}{9}

\bibitem{Dittmar}  A.~De Roeck {\it et al.},
  Eur.\ Phys.\ J.\  C {\bf 66}, 525 (2010)

\bibitem{PDF_WG} R. Placakyte, S. Alekhin, D. Colferai and J.W. Huston, summary talk at
 "XVIII International Workshop on Deep-Inelastic Scattering and Related Subjects April 19 -23, 2010, Firenze, Italy"

\bibitem{GuiMou}  M.~Guidal and H.~Moutarde,
  Eur.\ Phys.\ J.\  A {\bf 42}, 71 (2009)
 
\bibitem{MulTan} P.~J.~Mulders and R.~D.~Tangerman,
   Nucl.\ Phys.\  B {\bf 461}, 197 (1996)
  [Erratum-ibid.\  B {\bf 484}, 538 (1997)]
 
\bibitem{BelRad} A.~V.~Belitsky and A.~V.~Radyushkin,
  Phys.\ Rept.\  {\bf 418}, 1 (2005)
   
\bibitem{NNPDF} R.~D.~Ball, L.~Del Debbio, S.~Forte, A.~Guffanti, J.~I.~Latorre, J.~Rojo and M.~Ubiali,
  Nucl.\ Phys.\  B {\bf 838}, 136 (2010)
    
\bibitem{Carnahan} H. Honkanen, S. Liuti, J. Carnahan, Y. Loitiere, P. R. Reynolds, Phys. Rev. D 79, 034022 (2009)

\bibitem{Perry_dis10} D.~Z.~Perry, K.~Holcomb and S.~Liuti,
  arXiv:1008.2137 [hep-ph].
  
 
\bibitem{Kohonen} T. Kohonen, Self-organizing Maps (Springer, New York, 2001), 3rd. ed.

\bibitem{progress} K. Holcomb, S.Liuti, D.Z. Perry, and S.K. Taneja, {\it in preparation}.

\bibitem{CTEQ6} J. Pumplin, D. R. Stump, J. Huston, H. L. Lai, P. Nadolsky, W. K. Tung (CTEQ6), Phys.\ Rev.\  D {\bf 65}, 014013 (2001)
 
\bibitem{AMP06} S. Alekhin, K. Melnikov, F. Petriello (AMP06), Phys. Rev. D 74, 054033 (2006) 


\end{thebibliography}
\end{document}